# Paramagnetic tunneling state concept of the magnetic anomalies of amorphous insulators

Alexander Borisenko and Alexander Bakai
*National Science Center "Kharkiv Institute of Physics&Technology", 1 Akademichna str., 61108 Kharkiv, Ukraine*

A generalized tunneling model of insulating glasses is formulated, considering tunneling states to be paramagnetic centers of spin 1/2. The expression for magnetic field dependent contribution into the free energy is obtained. The derivation is made of the expression for the nonmonotonic magnetic field dependence of dielectric constant, recently observed in several glasses in the millikelvin temperature range.

In the past few years a puzzling behavior of some multicomponent insulating glasses in magnetic field at super low temperatures was discovered. For example, the dielectric constant $\varepsilon$ of amorphous BaO-$Al_2O_3$-$SiO_2$ (BAS) in complex nonmonotonic manner depends on the magnetic field at T≤50 mK [1]. A sharp kink in dielectric constant of this material is observed, indicating a phase transition at T=5.84mK [2]. Within the millikelvin temperature range BAS also demonstrates the pronounced dependence of spontaneous polarization echo amplitude on magnetic field [3]. Similar results are reported for another multicomponent glass BK7 [4]. The heat capacity of these materials also depends on magnetic field in a nonmonotonic manner (see [5] and references therein).

It should be mentioned that the magnitudes of these anomalous effects don't simply scale with the concentration of magnetic impurities in the samples [4], thus allowing to rule out their direct impact. Neither these properties can be interpreted as a magnetoeffect, characteristic to nonlinear dielectrics, where the quadratic field dependence of $\varepsilon$ should be found [6].

The obvious conclusion from the above facts is that some glasses posses a subsystem that is susceptible to magnetic fields and that this subsystem is related to the structural features of amorphous insulators.

At zero magnetic field the anomalous physical properties of vitreous insulators at $T \le 1K$ are successfully described by the tunneling model (see [7] and references therein).

In the simplest case a tunneling state (TS) can be considered as an effective particle confined in a double well (W) potential. In this case due to overlap of the ground state wavefunctions in the two wells the ground energy level splits into a doublet with a gap $E = \sqrt{h^2 + \Delta^2}$ , $\Delta$ being the gap value for symmetric potential, $h$ being the difference of ground state energies in two wells neglecting tunneling. Any other excited states of this system, except this doublet, are neglected. The parameters $h$, $\Delta$ are commonly assumed to be random, obeying the phenomenological distribution:

$$P(h,\Delta) = P_0/\Delta\,,\; E_{\min} \le E(h,\Delta) \le E_{\max} \qquad (1)$$

The phenomenological parameters of the distribution function (1) are the constant $P_0$, proportional to the volume density of TSs and the lower and upper energy cutoffs $E_{\min}$ and $E_{\max}$. Although one has no reason to believe that the distribution function has so simple form for all substances in the wide region of parameters and temperatures, it can be used as a trial, properly parametrized function.

It is natural to assume that TSs are responsible for the mentioned above puzzling magnetic properties too. The models elaborated up to now may be separated into "orbital" and "spin" group according to the assumed origin of TS magnetic moment.

In the orbital models, tunneling of an electrically charged particle between potential minima can occur along different paths, so the presence of a magnetic field yields an Aharonov-Bohm phase and a change of energy eigenvalues. In a recent publication [6] a hat-shaped W potential was considered with two minima in the azimutal direction along the rim. Within this model, the experimentally observed on BAS samples maximum in the real part of dielectric constant at $B \approx 0.1T$ requires one to assume the TS electric charge to be $Q \propto 10^5 |e|$, where $e$ is an elementary charge. The authors speculate the origin of such a large value of $Q$ to result from the strong cooperative interaction between TSs at low temperatures, when the quasiparticles should be considered rather then the "bare" TSs.

Instead of a hat-like potential a multi-well potential may be considered [5]. Qualitatively similar results are obtained within the framework of this model and again a big value of $Q$ is needed to fit the experimental data for the heat capacity of BAS and Duran glasses.

The multi-path tunneling may be realized in a different way, considering interaction between pairs of TSs of certain relative orientation and closed tunneling sequences in this complex [8]. In the presence of magnetic field this system possesses two orbital quantum states with a linear field dependence of the energy levels. Under the experimental conditions, the deduced dielectric susceptibility shows an oscillatory behavior, with an effective flux quantum of the order of $10^{-5}\ \hbar/e$ . However, this model is formulated for the near-degenerate TSs $(h \to 0)$ and, as we are concerned, it hasn't been developed for the case of general W potential asymmetry up to now.

In the models of the "spin" group the intrinsic magnetic moments associated with the tunneling entity are considered. The model by Bodea and Wurger [9] considers nuclear origin of spins of TSs. In a general case the nucleus quadrupole momentum is not zero and depends on the spin projection $\hat{I}_z$. The tunneling motion is then coupled to Zeeman energy due to the inhomogeneity of electric molecular field. This leads to the quadratic magnetic field contribution in $\varepsilon$. However, this model fails to predict the nonmonotonic field dependence of $\varepsilon$ and the amplitude of the effect is by several orders of magnitude smaller then the experimentally observed at $B \sim 1T$.

In this communication we consider a magnetic field effect on TS ensemble assuming them to be paramagnetic centers. There are different possible physical mechanisms of TS paramagnetism origin in covalent solids, for example:

i) nuclear paramagnetism of a tunneling atom;

ii) paramagnetism of the electrons or holes, localized on unsaturated ('dangling') bonds [10] in the vicinity of a W potential represented by a split atomic site occupied by tunneling atom. Impurity atoms, present in the sample, can act as donors of such electrons and holes;

iii) paramagnetism due to Van Vleck mechanism [11].

Any of these mechanisms or their combinations can be relevant for a given case.

We shall find the contribution of noninteracting spin 1/2 paramagnetic tunneling state (PTS) into the free energy of an amorphous solid at nonzero magnetic field. The generalization of this model to the case of higher spins is straightforward. It is assumed that tunneling between two wells does not conserve the spin projection. It will be shown how PTSs could give rise to nonmonotonic magnetic field dependence of thermodynamic values.

We write down the operators of two-state coordinate (l, r) and spin projection (u, d) variables in the representation of Pauli matrices $\sigma_z$ and $\tau_z$ respectively.

The tunneling transition operator between states with equal spin projection is then represented by the Pauli matrix $\sigma_x$. The frequency of this tunneling motion is taken to be $\Delta/\hbar$.

The tunneling transitions in which both the spin and coordinate states are changing simultaneously, are convenient to describe in terms of creation and annihilation operators:

$\sigma^+ = (\sigma_x + i\sigma_y)/2$; $\sigma = (\sigma_x - i\sigma_y)/2$;

$\tau^+ = (\tau_x + i\tau_y)/2$; $\tau = (\tau_x - i\tau_y)/2$,

which act in the following way:

$\sigma^+|l\rangle = |r\rangle$; $\sigma^+|r\rangle = 0$; $\sigma|l\rangle = 0$; $\sigma|r\rangle = |l\rangle$;

$\tau^+|u\rangle = |d\rangle$; $\tau^+|d\rangle = 0$; $\tau|u\rangle = 0$; $\tau|d\rangle = |u\rangle$.

We take the frequency $D/\hbar$ for the tunneling transitions $|ru\rangle \leftrightarrow |ld\rangle$, governed by the operator $\sigma\tau^+ + \sigma^+\tau$. The tunneling transitions $|lu\rangle \leftrightarrow |rd\rangle$, governed by the operator $\sigma^+\tau^+ + \sigma\tau$, may be deduced from the former scheme by inversion of time $(t \to -t)$ and hence the frequency $-D/\hbar$ is associated with them.

Note that the variables (l, r) and (u, d) are considered to be independent, and hence $\sigma_i$ and $\tau_j$ commute with each other.

In these notations, we can write down the single PTS Hamiltonian as follows:

$$\hat{H} = -1/2 \cdot \left(h\sigma_z + \Delta\sigma_x + u\tau_z + D\sigma_y\tau_y\right). \qquad (2)$$

Here $h$ is the difference of ground states of two wells, $u$ is a Zeeman energy splitting due to external magnetic field:

$$u = g\mu_{PTS}B, \qquad (3)$$

$g$ is the Lande factor, $\mu_{PTS}$ is the PTS magneton, $B$ is the absolute value of the external magnetic field.

We assume direct coupling of the potential asymmetry $h$ to electric field **E** through the PTS intrinsic electric dipole moment **p**:

$$h = h_0 + 2(\mathbf{p} \cdot \mathbf{E}). \qquad (4)$$

Writing $\hat{H}$ in the basis $\{|lu\rangle, |ru\rangle, |ld\rangle, |rd\rangle\}$ and diagonalizing the resulting 4x4 matrix, we obtain the energy spectrum:

$$E_{S,A} = \mp\frac{1}{2}\sqrt{G_+^{\ 2} + D^2};\ E_{2,3} = \pm\frac{1}{2}\sqrt{G_-^{\ 2} + D^2} \qquad (5)$$

and the normalized eigenvectors:

$$\psi_S = \frac{1}{2}\begin{pmatrix}1\\1\\1\\1\end{pmatrix},\ \psi_A = \frac{1}{2}\begin{pmatrix}-1\\1\\1\\-1\end{pmatrix},\ \psi_{2,3} = \frac{1}{2}\begin{pmatrix}-1\\\pm1\\\mp1\\1\end{pmatrix}. \qquad (6)$$

In the equation (5) the abbreviations $G_\pm \equiv u \pm \sqrt{h^2 + \Delta^2}$ are used.

As expected, the symmetric state $\psi_S$ has the lowest energy, the antisymmetric state $\psi_A$ has the highest energy, while the mixed states $\psi_{2,3}$ form the internal doublet.

Note that the parameter $D$ is responsible for the nonmonotonic magnetic field dependence of the eigenvalues $E_{2,3}$. This feature is essential in our model to describe the nonmonotonic magnetic field dependence of PTS contribution into thermodynamic quantities.

Using the expression for single PTS energy spectrum (5) and putting the Boltzmann constant $k_B = 1$, one can obtain an expression for single PTS free energy $f = -T\ln Sp\exp(-\hat{H}/T)$ in the explicit form:

$$f = -T\ln\left\{2\cosh\frac{\sqrt{G_+^{\ 2} + D^2}}{2T} + 2\cosh\frac{\sqrt{G_-^{\ 2} + D^2}}{2T}\right\}. \qquad (7)$$

With $u, D \to 0$, expression (7) takes the form of conventional single TS free energy with additional term $-T\ln 2$, originating from unresolved magnetic doublet.

The thermodynamic variable of our special interest is the dielectric susceptibility. For the static case we have:

$$\chi_{\alpha\beta}(\omega=0)=-\frac{1}{V}\left(\frac{\partial^2 f}{\partial E_\alpha \partial E_\beta}\right)_T , \qquad (8)$$

where **E** is an external dc electric field, $V$ is the sample volume.

As the relaxation time scale should become large in the temperature range of our interest, $T \le 100mK$ [12, 13], the relaxation effects have to be considered while comparing our results with experimental data.

At low (on the energy gap scale) frequencies the expression for PTS ac dielectric susceptibility is as follows:

$$\chi(\omega)=\chi_{res}+\frac{\chi_{rel}}{1-i\omega\tau} , \qquad (9)$$

where $\chi_{res}$ and $\chi_{rel}$ are the resonant and relaxation parts respectively, $\omega$ is an external field frequency, $\tau$ is a relaxation time.

The explicit form of expressions for $\chi_{res}$ and $\chi_{rel}$ is a bit cumbersome to display here. It comes out that all the terms in (8) proportional to $T^{-1}$ concern the relaxation component, while the rest contribute to the resonant part.

Both $\chi_{res}$ and $\chi_{rel}$ contain the terms which depend nonmonotonically on Zeeman splitting, for nonzero $\sqrt{h^2+\Delta^2}$ . We denote them $\chi_{res}^{nm}$ and $\chi_{rel}^{nm}$ respectively:

$$\chi_{res}^{nm} \sim 2\sinh\frac{\sqrt{G_-^{\,2}+D^2}}{2T}\frac{h^2D^2}{\left(h^2+\Delta^2\right)\left[G_-^{\,2}+D^2\right]^{3/2}} \qquad (10)$$

$$\chi_{rel}^{nm} \sim \cosh\frac{\sqrt{G_-^{\,2}+D^2}}{2T}\frac{G_-^{\,2}h^2}{\left[G_-^{\,2}+D^2\right]\left(h^2+\Delta^2\right)T} \qquad (11)$$

At $u=\sqrt{h^2+\Delta^2}$ $\chi_{res}^{nm}$ has a maximum, while $\chi_{rel}^{nm}$ has a minimum, which cancel each other for $D << T$ . The widths of these nonmonotonic regions are proportional to $D$.

Numerical calculations of PTS dielectric constant show that the extrema shapes and positions strongly depend on the PTS density of states.

For illustration of our results we present calculations of the PTS ensemble dielectric constant, integrated over the phenomenological density of states (1) and averaged over the isotropic distribution of dipole moments with the fixed absolute value $p_0$.

The chosen values of model parameters are shown in Table I. They are compatible with the data known from literature (see e. g. [13]).

TABLE I. The values of fitting parameters.

| $4\pi P_0 p_0^{\,2}/3\varepsilon_0$ | $E_{\min}, K$ | $E_{\max}, K$ | $\Delta/D$ |
|---|---|---|---|
| $1.5\cdot 10^{-4}$ | $10^{-4}$ | 1 | 1 |

The value of parameter $\Delta/D=1$ in Table I corresponds to the equal probabilities of spin projection flip and conservation during tunneling.

In Figs.1, 2 we present the dependence of the real $(\varepsilon')$ and imaginary $(\varepsilon'')$ parts of PTS ensemble dielectric constant on Zeeman splitting $u$ at two different temperatures. For the rough check of relaxation effects we consider different values of the parameter $\omega\tau$ and neglect the dependence of relaxation time $\tau$ on the parameters $(h,\Delta)$ and $u$.

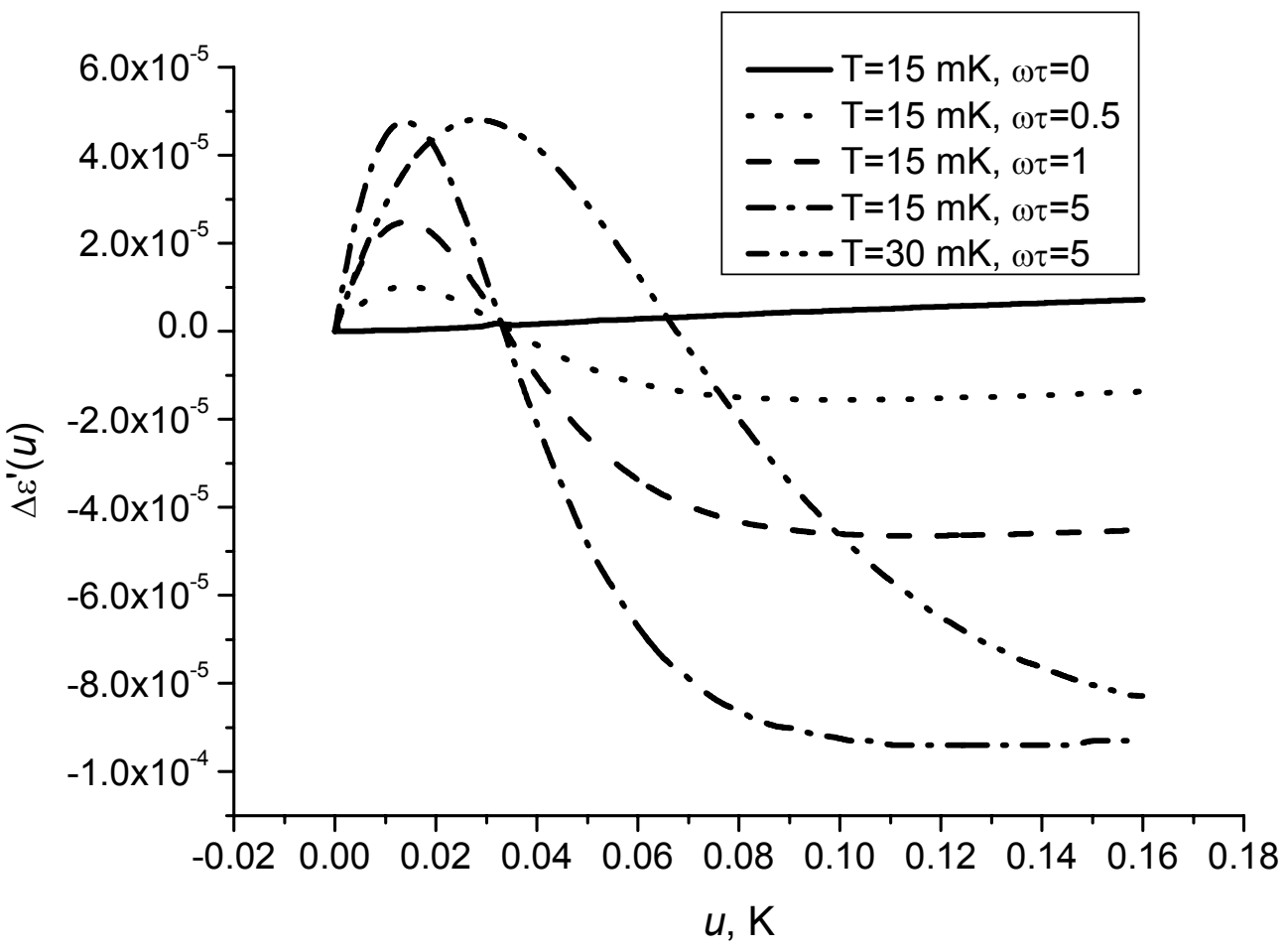

FIG.1. The real part of the PTS ensemble dielectric constant vs. Zeeman energy splitting $(\Delta\varepsilon'(u)=\varepsilon'(u)-\varepsilon'(0))$.

In the adiabatic regime $(\omega\tau=0)$ no nonmonotonic dependence of $\Delta\varepsilon'(u)$ in Fig. 1 is observed since the nonmonotonic parts of the resonant and relaxation dielectric susceptibility completely cancel each other (see (10), (11)). The amplitude of variation of $\Delta\varepsilon'(u)$ depends on the parameter $\omega\tau$ roughly as $1-\left(1+\omega^2\tau^2\right)^{-1}$ (see (9)).

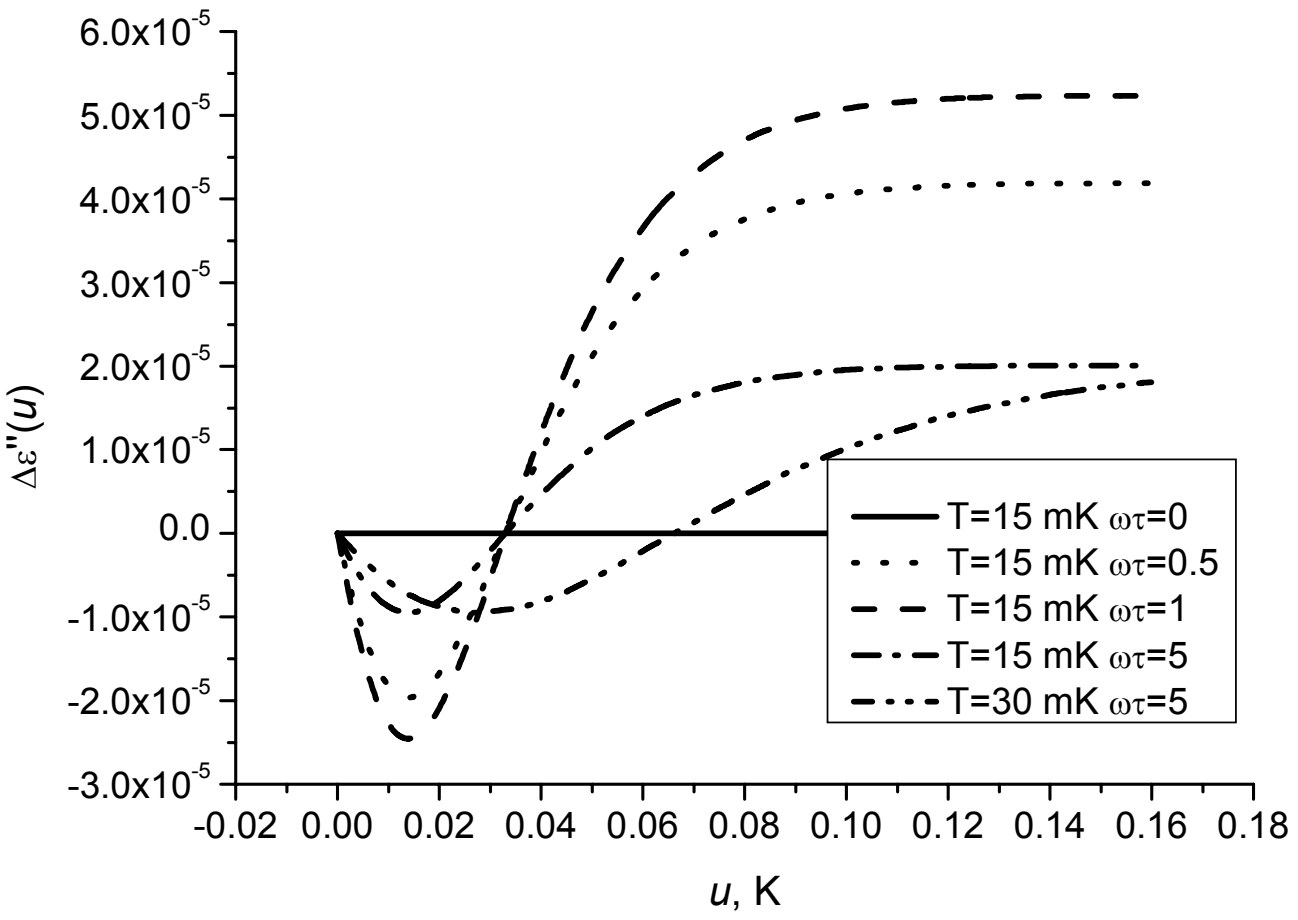

FIG.2. The imaginary part of PTS ensemble dielectric constant vs. Zeeman energy splitting $(\Delta\varepsilon''(u)=\varepsilon''(u)-\varepsilon''(0))$.

The amplitude of variation of $\Delta\varepsilon''(u)$ in Fig. 2 depends on the parameter $\omega\tau$ in a nonmonotonic manner $\omega\tau\left(1+\omega^2\tau^2\right)^{-1}$ (see (9)). Both in the adiabatic regime $\omega\tau = 0$ and in the strong relaxation regime $\omega\tau \to \infty$ $\Delta\varepsilon''(u)$ becomes flat.

Comparing the curves for two different temperatures in Figs. 1, 2, one can see that the positions of extrema on both plots are close to the value T, which is natural since there is no extra appropriate energy parameter left after integration over the smooth distribution function (1). This is a characteristic feature of the chosen density of states (1).

We also calculated the dependence of the PTS heat capacity $c_v = -T\dfrac{\partial^2 f}{\partial T^2}$, integrated over the distribution function (1), on Zeeman splitting and found that it has a maximum which position scales with temperature too. The dependence of amplitude of spontaneous polarization echo, calculated in the framework of PTS model, on $u$, reproduces the essential features of experimental data [3]. These results are to be published elsewhere.

The dielectric measurements on BK7 glass in the temperature range $15mK \le T \le 52mK$ show the pronounced maximum of $\varepsilon'(B)$ at $B \propto 10T$. At the same strength of the magnetic field the minimum in dielectric loss tangent $(\sim \varepsilon'')$ is observed. This feature is qualitatively reproduced in the frame of our model, giving the correct orders of magnitude of the extrema heights for the reasonable choice of the free fitting parameters (see Table 1) and the value of relaxation parameter $\omega\tau$ of order of unity. The experimentally measured magnitudes of both extrema decrease with the increase of the temperature. We suppose this to occur due to the temperature dependence of relaxation time $\tau$ [13]. Our calculations reveal the scaling of extremum field strength with the temperature. Though the experimental extrema positions slightly depend on the temperature, this result doesn't seem to be true. The reason may be that the real distribution is essentially different from that given by the trial expression (1). To avoid this discrepancy, a modified density distribution may be used instead of (1), for example the one with a step in the low energy region [1].

The minimum of $\varepsilon'(B)$ and the maximum of the dielectric loss of BK7 at $0.1T \le B \le 1T$ are not reproduced in the frame of our model, at least for the given TS density of states (1).

The experimental curves for BAS also show a distinct maximum in $\varepsilon'(B)$ at $B_{\max} \approx 0.1T$, though the dielectric loss for this material doesn't seem to have a distinct minimum at this field value. From the calculated value $u_{\max} \approx 15mK$ and the experimentally measured extremum magnetic field $B_{\max} \approx 10T$ at T=15 mK one can deduce the value of PTS magneton for BK7: $\mu_{PTS}(BK7) \propto 10^{-26} J/T$, which is "close" to the value of nucleus magneton $\mu_n = 5\cdot 10^{-27} J/T$. The experimentally measured maximum in $\varepsilon'(B)$ for BAS at the same temperature at $B_{\max} \approx 0.1T$ gives $\mu_{PTS}(BAS) \propto 10^{-24} J/T$, which is "close" to the value of Bohr magneton $\mu_B = 9.27\cdot 10^{-24} J/T$. We remind once more that these calculations are based on the trial distribution function (1), so the results are of the qualitative character.

Such a big difference in PTS magnetons in these materials may be possibly attributed to the big difference of magnetic impurities concentration (about 100 ppm and less then 6 ppm for BAS and BK7 respectively [4]). As it was mentioned above, TSs can be localization centers for impurity electrons and (or) holes.

In summary, we propose a PTS model for amorphous solids, which predicts the magnetic field dependence of thermodynamic quantities. In particular, we calculate the dielectric constant of PTS ensemble for the case of spin 1/2 and find out that our results reproduce the general features of experimental data for BK7 and BaO-$Al_2O_3$-$SiO_2$ glasses [1, 4]. We suppose that PTS model may be adequate for description of other disordered solids, e. g. crystals with off-center impurities.

[1] P. Strehlow, M. Wohlfahrt, A. G. M. Jansen, R. Haueisen, G. Weiss, C. Enss and S. Hunklinger, Phys. Rev. Lett. 84, 1938 (2000).

[2] P. Strehlow, C. Enss and S. Hunklinger, Phys. Rev. Lett. 80, 5361 (1998); S. Hunklinger, C. Enss and P. Strehlow, Physica B 263-264, 248 (1999).

[3] S. Ludwig, C. Enss, P. Strehlow and S. Hunklinger, Phys. Rev. Lett. 88, 075501-1 (2002).

[4] M. Wohlfahrt, P. Strehlow C. Enss and S. Hunklinger, Europhys. Lett. 56 (5), 690 (2001))

[5] G. Jug, Philos. Mag. 84, 3599 (2004)

[6] S. Kettemann, P. Fulde and P. Strehlow, Phys. Rev. Lett. 83, 4325 (1999)

[7] P. Esquinazi (ed.) *Tunneling Systems in Amorphous and Crystalline Solids* (Springer, Berlin, 1998).

[8] A. Wurger, Phys. Rev. Lett. 88, 075502-1 (2002).

[9] D. Bodea, A. Wurger, J. Low Temp. Phys. 136, 39 (2004).

[10] I. Solomon, in *Amorphous Semiconductors*, Topics in Applied Physics Vol. 36, edited by M.H. Brodsky (Springer Verlag, Berlin – Heidelberg - New York, 1979).

[11] J. H. Van Vleck, *The Theory of Electric and Magnetic Susceptibilities* (Oxford, 1932).

[12] S. Rogge, D. Natelson and D. D. Osheroff, J. Low Temp. Phys. 106, 717 (1997).

[13] A. Burin, J. Low Temp. Phys. 100, 309 (1995).